\documentclass[amsmath,amssymb,iopams]{iopart}

\usepackage{graphicx}

\begin{document}

\title{Optical conductivity of a Hubbard ring with an impurity}
\author{Cosima Schuster and Philipp Brune}
\address{Institut f\"ur Physik, Universit\"at Augsburg, 86135 Augsburg, Germany}
\ead{cosima.schuster@physik.uni-augsburg.de}
\date{\today}

\begin{abstract}
We investigate the optical conductivity of a
               Hubbard ring in presence of an impurity
               by means of exact diagonalization using the
              Lanczos algorithm. We
               concentrate thereby on the first excited,
               open shell state, i.e.\ on twisted 
               boundary conditions. In the metallic phase a
               substantial part of the spectral weight lies in
               the Drude peak, $\sigma(\omega) = D_c\delta(\omega)+\sigma_{\rm reg}$.
 In the non-interacting system,  the Drude peak can be visualized in our calculations
               at $\omega = 0$ even for finite chain lengths. 
               Adding the impurity, 
               the main peak is shifted
                to finite frequencies proportional to the impurity strength. The shift
indicates the energy gap of the disturbed finite size system, 
also in the interacting system. 
Thus, we can pursue in the optical conductivity for finite metallic systems  the
energy gap.
However, due to level crossing, the impurity-induced peak
arises in the interacting system first when a certain impurity strength is exceeded.
In the Mott insulating phase, the impurity  leads to an impurity state within the gap. 
\end{abstract}

\section{Introduction} 
The optical properties of strongly interacting electron systems are of great interest,
primarily to understand the properties of metal-insulator transitions \cite{Carmelo,Fehske}.
We consider here the optical properties of Hubbard rings.
The attention is directed to one-dimensional systems, 
because the investigation of low dimensional systems has already provided important
 insights \cite{Mattis-Buch,Shankar}.
One-dimensional systems are studied extensively 
both experimentally and theoretically in the context of correlated systems \cite{Schwerpunkt}.
In addition, they are of great interest in the context of disordered systems \cite{Kramer-rep}.
Furthermore, powerful theoretical tools exist for the theoretical treatment of 
one-dimensional systems, like bosonization \cite{Haldane81}, Bethe-ansatz \cite{Korepin}, 
conformal field theory \cite{Affleck}, and the
density matrix renormalization group \cite{dmrg}. Also exact diagonalization is more powerful 
in low dimensions since larger systems can be investigated.

In order to study the localization due to disorder in the one-dimensional Hubbard-model with
regard to a metal-insulator transition,
we  investigate the optical conductivity of a Hubbard ring with impurity.
In our work we ask the following questions: 
 How do the  different energy scales  -- 
$v_F/L$, $\epsilon$, $\Delta$ -- interplay in the metallic phases?
Here $v_F$ is the Fermi velocity, $L$ the system size, $\epsilon$ the characteristic local
impurity strength, and $\Delta$ the energy gap.
How does the energy gap in the insulating phase, and the absorption edge seen in the 
optical conductivity change with the impurity?
Especially, we concentrate on a weak  impurity and weak interaction.
The transition from a Mott-insulator to an Anderson-insulator  was studied previously
in case of strong disorder and strong interaction
\cite{Dietmar,Rudo}. 

The outline of the article is as follows.
First we discuss the known properties of the disordered Hubbard-model.
Then we describe our numerical approach to the optical conductivity.
In section 4 we present our numerical results.
\section{The Hubbard model}
\subsection{The clean model}
The Hubbard model is the generally accepted prototypical model
to describe the interplay between kinetic energy ($\to$ delocalization) and
local interaction ($\to$ localization) for electronic systems.
The Hubbard chain is, in addition, exactly solvable by means of the Bethe ansatz \cite{LiebWu}.
The Hamiltonian of the Hubbard model is given by
\begin{equation}\label{Hhub}
H_{\rm Hubb}= -t\sum^{N}_{i,\sigma}
 \left( c^+_{i,\sigma}c^{}_{i+1,\sigma} + {\rm h.~c.}\right)
   + U\sum^{N}_{i}n_{i,\uparrow}n_{i,\downarrow}.
\end{equation}
The chain length is given by  $L=Na$, $a$ is the lattice spacing, and
  $n_0=N_e/N$ is the particle density. In addition, we define
$n_i=n_{i\uparrow}+n_{i\downarrow}$, $m_i=n_{i\uparrow}-n_{i\downarrow}$,
and the magnetization $M=\sum_i\langle S^z_i\rangle=N_\uparrow-N_\downarrow$.
In the clean case,
the model shows three phases. 
        Phase one, the Luttinger liquid phase, arises for $U\geq 0$ and away from half filling. Spin- and
charge-excitations are those of a Luttinger liquid.
The point $U=0$, $n=1$ (non-interacting electrons and half filling) also belongs to this phase.
Phase two,  which is called spin gap phase, 
occurs for $U<0$, where the spin-excitation spectrum has a gap and the
low-lying charge-excitations can be described by those of a Luttinger
        liquid \cite{Haldane81}. 
The last phase, the Mott insulating phase,  occurs for
$U>0$ and half filling, where the charge excitations have a gap and the
spin-excitations are those of a Luttinger liquid.
In the clean case, the Hubbard model shows
 spin-charge separation which is characteristic for Luttinger liquids
\cite{Yacoby,Claessen}.
A longer ranged interaction,
even if it is the more generic case,
leads to additional ground state phases \cite{Voit}.
\subsection{Impurities in interacting Fermi systems}
We introduce disorder in our model by adding local potentials $\epsilon_i$ 
to the above Hamiltonian, Eq.\ (\ref{Hhub}),
\begin{equation}\label{Hhube}
H=H_{\rm Hubb} +\sum^{N}_{i}\epsilon_i n_i\; .
       \end{equation}
The most studied example of impurity effects is 
 an interacting system of spinless fermions in the presence of a local  potential
scatterer of strength $\epsilon$, i.e.\ $\epsilon_i=\epsilon\delta_{i,0}$. The behavior of this system is well  known.
In this case, the free motion of the fermions inside the ring is restricted
mainly due to the backscattering  at the impurity, 
$\pm k_F\to \mp k_F$.
As discussed by Kane and Fisher \cite{Kane92} the
impurity strength
scales to zero (becomes transparent) for an attractive
interaction and scales to infinity (becomes completely reflective) for a repulsive
interaction, according to the renormalization group equation
\begin{equation}\label{KF}
\frac{{\rm d}\epsilon}{{\rm d}l}=(1-K)\epsilon\;.
\end{equation}
The Luttinger parameter $K$ depends on the interaction such that $\infty>K>1$ for attractive interaction,
$K=1$ in the non-interaction system, and $1/2<K<1$ for repulsive interaction. In the charge ordered
insulating phase at half filling, which corresponds to the Mott phase of the Hubbard model, $K=0$.

In a Hubbard chain the impurity is accordingly  known to be relevant in the Luttinger liquid phase.
The scaling of the impurity \cite{Kane92},
\begin{equation}
\label{scaling}
\frac{{\rm d}\epsilon}{{\rm d} \ln L}=(2-K_c-K_s)\epsilon \;,
\end{equation}
contains now the  Luttinger parameters $K_c$ and $K_s$ for the charge and spin degrees of freedom.
As before,  $K_c=1$ and $K_s=1$ in the non-interacting system,  $K_c>1$ for attractive interaction
and $K_c<1$ for repulsive interaction. $K_s=0$ in the spin gap phase $K_s=0$ and $K_s=1$
for repulsive interaction.
In contrast to spinless fermions, no metal-insulator transition as function of the interaction is found 
in the Hubbard-model in presence of the impurity; the system is always localized.
The scaling beyond Eq. (\ref{KF}) and (\ref{scaling}) is discussed by Meden et al. \cite{Meden}.
They  show that the scaling of the
               impurity -- as shown in Eq. (\ref{scaling}) and 
predicted by bosonization, is  only valid for
               long chains and strong interaction.

These results for  a single impurity can be extended on disordered systems \cite{Kramer-rep}.
Thereby   a metal-insulator transition is found in the spinless Fermi model at finite disorder 
strength for strong attractive interaction \cite{Doty}.
In the Hubbard model it is found that  a weak repulsive  interaction reduces the 
localizing effect of the disorder \cite{RomPun}. 
However, it is also known  \cite{mori} that the $4k_F$ contribution 
of the scattering potential,  not included in the above scaling relation, is responsible 
for the localization in interacting systems. 
The $4k_F$ contribution becomes relevant for strong interaction.
Whereas the $2k_F$ 
impurity-scattering and interaction weaken each other,
 thus leading to  weaker localization than in the non-interacting system,
the $4k_F$ contribution enhances the localization for strong interaction.

Many recent activities related  to the disordered Hubbard model concentrate on single defects. 
For example, the Friedel oscillations  are investigated 
               \cite{ Kramer-rep,B4G:Bed98,Schreiber-rep}.
The exponent of the algebraic decay depends strongly on interaction and resembles the scaling
of the impurity for a given interaction.
Thus,  studying the local behavior,
possibly the relevant theoretical model can be
identified \cite{Rommer}.              
Concerning the insulator-insulator transitions which occur in the disordered interacting system
(Anderson insulator versus Mott insulator) \cite{Dietmar,Vollha}
and the Peierls-Hubbard model (band insulator versus Mott insulator)  
\cite{Fehske,Jeckelmann00},
the Drude weight -- or the phase sensitivity -- and
 the optical conductivity were used to characterize the different insulating phases.

\section{Optical conductivity}
In the linear response approximation, the optical conductivity gives the response of a system to an external electric field, $\langle {\bf j}\rangle =\sigma{\bf E}$, ${\bf E}=-\dot{\bf A}/c$. 
 The Kubo formula is used to evaluate the optical conductivity
 $\sigma(t,t')$ by means of the current-current correlation function $<[{\bf j}(t),{\bf j}(t')]>$,
 \begin{equation}  \langle{\bf j}(t)\rangle+{\bf j}_{\rm dia}=i\int_{-\infty}^t {\rm d}t' \left \langle \left[{\bf j}(t),{\bf j}(t')\right]\right\rangle\frac{{\bf A}(t')}{c}-\frac{ne^2}{mc}{\bf A}(t) \;, 
\end{equation}
with ${\bf j}=e{\bf p}/m$.
Using the Green's functions formalism to evaluate the propagation of the electrons, 
the optical conductivity is given by
\begin{eqnarray}\label{green}
\sigma({\bf q},\omega)&=&\left\langle 0\left|{\bf j}_q^+\frac{1}{\omega+i\delta-H}{\bf j}_q\right|0\right\rangle \label{sigma}\\
{\rm Re}\sigma(\omega)&=& \sum_n|\langle 0|{\bf j}_q|n\rangle|^2\delta\big(\omega-(E_n-E_0)\big)\;.
\label{Resigma}
\end{eqnarray}
The current operator for a one-dimensional lattice model is given by
\begin{eqnarray}\label{jone}
j_l&=&\sum_{\sigma}(c^+_{l+1,\sigma}c_{l,\sigma} -c^+_{l,\sigma}c_{l+1,\sigma})\\
j_q&=&\sum_lj_l\exp(iql)=2i\exp(iq/2)\sum_{k\sigma}\sin(k+q/2)c^+_{k,\sigma}c_{k-q,\sigma}\;.
\end{eqnarray}
Using the fact that the Fermi surface of a chain consists of only the two Fermi points
at $\pm k_F$, the  current operator can be written in an intuitive
way, assuming linear dispersion relation near the Fermi points in the non-interacting system,
$\epsilon(k)\propto v_Fk$, $v_F=-2t\sin(k_F)$,
\begin{equation}j=v_F\sum_k(n_{R,k}-n_{L,k})\label{mattisj}\end{equation}
where L, R denote left and right moving particles.
Since $j$ commutes with the kinetic and interaction part of the Hubbard Hamiltonian,
it is a good quantum number.
Using Eq. (\ref{jone})
the optical conductivity, Eqs. (\ref{sigma}) and (\ref{Resigma}) can be rewritten as
\begin{eqnarray}
{\rm Im}\sigma(\omega)&=&\frac{e^2}{\omega L}\left[-\langle H_{\rm kin}\rangle-{\cal P}
\sum_{n \neq 0}\frac{2\langle 0|j|n\rangle}{(E_n-E_0)-\omega^2}\right]\\
{\rm Re}\sigma(\omega)&=&D\delta(\omega)+\sigma_{\rm reg}\;.
\end{eqnarray}
The Drude weight $D$ is given by
$D=2v_cK_c$ in the Hubbard model, where $v_c$ and $K_c$ are 
interaction dependent constants which can be determined by means of the Bethe ansatz
\cite{RomPun,Schulz90}. The Drude weight is also related to the charge stiffness and the phase sensitivity \cite{Shastry}.  
Using  Eq. (\ref{mattisj}), first the conjecture of
persistent currents and infinite  optical conductivity was given \cite{Mattis}.
Finally, the optical conductivity obeys a sum rule,
\begin{equation}
\int_{\infty}^{\infty} {\rm Re}\sigma(\omega)d\omega=
-\frac{\pi e^2}{L}\langle H_{\rm kin}\rangle\;.
\end{equation}

Within the  Lanczos algorithm, also called Lanczos vector method \cite{Lanczos},  
the coefficients arising in the
representation of the correlation function  as a continued fractions 
can be determined in an exact
diagonalization calculation.
Extensions of this method to longer chains using  the
density matrix renormalization group treatment are discussed in \cite{Hallberg} and \cite{Kuehner}. They also give a comprehensive overview of the  Lanczos algorithm.
The Green's function in Eq. (\ref{green}) can be rewritten in continued fractions:
\begin{equation}
\sigma(\omega+i\delta)=\frac{\langle 0|j_q^+j_q|0\rangle}{\omega+i\delta-a_0-\displaystyle\frac{b_1^2
}{\omega+i\delta-a_1-\frac{b_2^2}{\omega+i\delta-\ldots}}}
\end{equation}
The coefficients $a_n$ and $b_n$ can be evaluated from
$\mu_n=\langle j_q^+H^nj_q\rangle$. 
Using a projective technique \cite{gagliano} the coefficients are determined
by the following recursion formulas:
\begin{eqnarray}
|f_0\rangle&=&j|0\rangle \\
|f_{n+1}\rangle&=&(H-a_n)|f_n\rangle-b_n^2|f_{n-1}\rangle\\
a_n&=&\langle f_n|j|f_n\rangle/\langle f_n|f_n\rangle\\
b_n&=&\langle f_n|j|f_n\rangle /\langle f_{n-1}|f_{n-1}\rangle\;.
\end{eqnarray}
\section{Numerical results}
\begin{figure}[hb]
\includegraphics[width=0.75\textwidth]{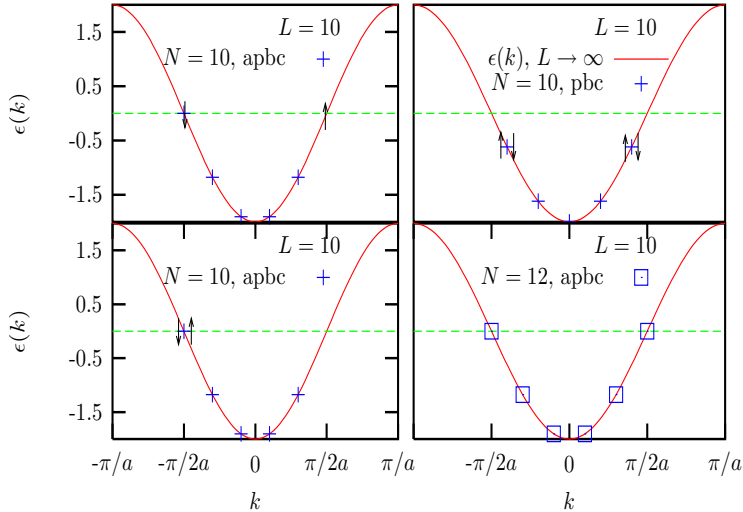}
\caption{Single particle ground state energy, $\epsilon(k)$, versus momentum, $k$, 
for periodic -- $k=2n\pi/L$  -- and anti-periodic 
-- $k= (2n+1)\pi /L$ -- boundary conditions. We compare the ground state with
ten and twelve electrons on  ten sites.  The true ground state for ten electrons is plotted in the upper right corner, the true ground state for twelve electrons in the lower right corner. 
The spins are only drawn for the levels near the 
Fermi energy, and   
the plus and the small square correspond to $\uparrow\downarrow$.
For the partly filled shell state -- upper left corner -- only one of the spin-degenerate states is shown. For the open shell states -- lower left corner -- only one realization is shown. As indicated in the upper right corner, the line gives the dispersion, $\varepsilon=-2t\cos k$
of the free tight binding model in the thermodynamic limit.}
\label{fig-rand}
\end{figure}

In the following, we present our numerical results, 
obtained for a half- or third-filled band,  in the 
four different sectors of the Hubbard model. Furthermore, we assume $M=0$, i.e.\ 
$N_{\downarrow}=N_{\uparrow}$.
We compare our results from the optical conductivity
with the energy gap, 
\begin{equation}\label{gap-eq}
\Delta(L)=[E(L,N_e+1)+E(L,N_e-1)-2E(N_e)]/2.
\end{equation}
In the non-interacting case we can visualize the Drude peak at $\omega=0$ even 
for finite system size.
In case of $N_e/2$ odd and anti-periodic boundary conditions
($c_{L+1}=-c_{1}$) or $N_e/2$ even and periodic boundary conditions ($c_{L+1}=c_{1}$)
 four degenerate states  contribute 
to the ground state. 
Due to this degeneracy of the partly filled and 
open shell states dissipationless excitations exist. 
Nevertheless, the true -- i.e.\ the state with the lowest energy --
ground state of a finite size system 
is obtained for  $N_e/2$ odd and periodic boundary conditions or vice versa.
The different configurations are shown in Fig. \ref{fig-rand}.
The upper left corner shows the partly filled shell state with a single particle at the Fermi points, the lower left corner shows the open shell state with two particles at one of the Fermi points,
the upper right corner shows the filled shell state according to the true ground state for
$N_e/2$ odd, and the lower right corner the  filled shell state according to the true ground state for
$N_e/2$ even.
The statement that the closed shell state is always the lowest  in energy  
holds also in the interacting model \cite{Shastry}, and in the disturbed
model \cite{Leggett}.

The twist in the boundary conditions is often modeled by a magnetic flux inside
the ring. Anti-periodic boundary conditions corresponds to $c_{L+1}=\exp({\rm i}\pi)c_{1}$ \cite{Shastry}.
Using this formulation, it is clear that the magnetic flux can induce a current in the system.
The  open shell states, which are realized for $N_e/2$ odd and anti-periodic
boundary conditions,  see the plot in the  bottom left corner of  Fig. \ref{fig-rand}, 
carry  momentum and current, see Eq. (\ref{mattisj}) and \cite{Mattis}. 
Therefore, they have  an overlap with the current operator.
In the following we show only results for boundary conditions according to the open shell states.
\subsection{$U=0$}
\begin{figure}[htb]
\includegraphics[width=0.7\textwidth]{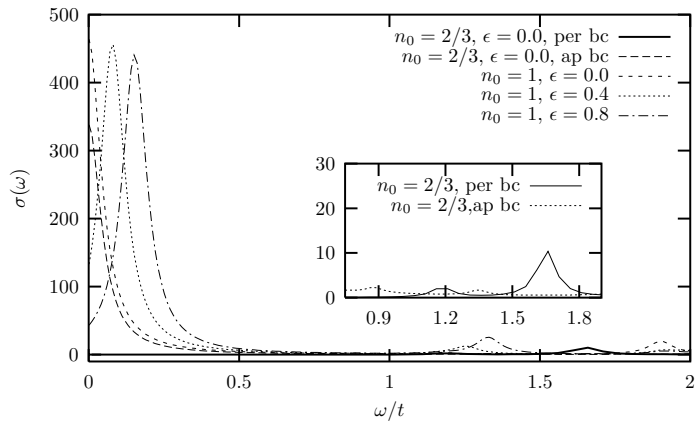}
\caption{Optical conductivity in arbitrary units versus frequency in units of the hopping rate $t$.
 System size and electron number
are $L=10$, $N_e=10$ in the
half-filled case and  $L=9$, $N_e=6$ in the third filled case. 
In the first case we compare different impurity strengths, in the latter case  we 
compare the ground state (periodic boundary conditions) with the
open shell state (anti-periodic boundary conditions). In the inset the higher excitations are shown for the third filled band.}
\label{fig-U0}
\end{figure}

As seen in Fig. \ref{fig-U0}, the Drude peak is found at $\omega=0$ in the clean system if we choose anti-periodic boundary conditions.  
According to $D=\pi v_F$,
the peak is higher in the half filled case than in the third filled case.
For the true ground state with the filled shell  no Drude peak or another 
prominent peak at higher frequencies 
with $\omega\approx 1/L$ is seen. The first peak at $\omega\approx 1.2t$ corresponds to 
the energy of  adding a particle above the filled shell.
Adding the impurity the energy levels of the non-interacting system 
 are partly shifted to higher energies. 

In case of  four lattice sites, the energies can be evaluated 
analytically. 
The open shell state with  $\varepsilon=0$
splits off in $\varepsilon=0$ and  $\varepsilon\approx 2\epsilon/L$
for small $\epsilon<t$. 
This relation applies also to longer chains.
The impurity 
does not change the structure of the electron distribution in coordinate space. 
The difference in the reorganization of the electron distribution 
due to an impurity  using periodic or anti-periodic boundary conditions
can be seen -- for longer chains numerically -- in the local density.
In the ground state, i.e.\ for the filled shell, 
the electrons are equally distributed.
An additional impurity induces decaying Friedel oscillations. 
In contrast, the electron distribution
shows oscillating behavior  -- already without the impurity -- if the degenerate
open shell state is considered. 
Adding the impurity, the electron density at the impurity 
site is reduced and we see the Friedel oscillations in addition.
Looking at momentum space we see, that the lower level corresponds to
the asymmetric superposition of the wave functions at the $k$-points at the 
Fermi-level 
$|0\rangle=\psi_{-\pi/2}-\psi_{\pi/2}$. The higher level, $|1\rangle$ 
corresponds to the 
symmetric superposition. 
The current operator connects both states,
$\langle 0|j|1\rangle\neq 0$.
Thus, adding the impurity the main peak shifts to higher frequencies,
with $\omega_{\rm P}\propto 2\epsilon/L$.
With increasing impurity strength, the peak-height decreases slightly, with 
nearly constant width, thus the peak loses weight. 
 This weight occurs dominantly in the next peak around $\omega
 \approx 1.2t$.
 In the thermodynamic limit, $L\to \infty$, the peak position
 shifts to zero frequency,
 $\omega_P \sim 1/L \to 0$. 
\subsection{Luttinger phase}
In the following,  we want to discuss the interacting  system
regarding a third filled band. 
In the numerical treatment we consider six electrons on nine lattice sites. 
With interaction the ground state cannot be viewed within in the band scheme.
The degeneracy of the states,
which have contributed to the ground state of the non-interacting system, is 
lifted. The  splitting is proportional to $U/L$.
The ground state with anti-periodic boundary conditions is -- roughly speaking  -- 
given by the symmetric superposition of partly filled shell states where the
spins are in spin-singlet states. The state of the next level
has no overlap  with the ground state via the current operator. 

However, adding the repulsive impurity to the interacting system,
there is an interplay of interaction and impurity.
Due to the interaction the charge distribution tends to be homogeneous,
 whereas the repulsive impurity repels the electron from one site.
In the band picture, the partly filled shell states split off
 due to the impurity. 
If the level splitting is smaller than the splitting due to 
interaction, no rearrangement is expected.
Increasing the impurity strength a state similar to the 
non-interacting system with impurity  becomes more favorable.
In this case,  there is an overlap between the lowest states. 

\begin{figure}[htb]
\includegraphics[width=0.75\textwidth]{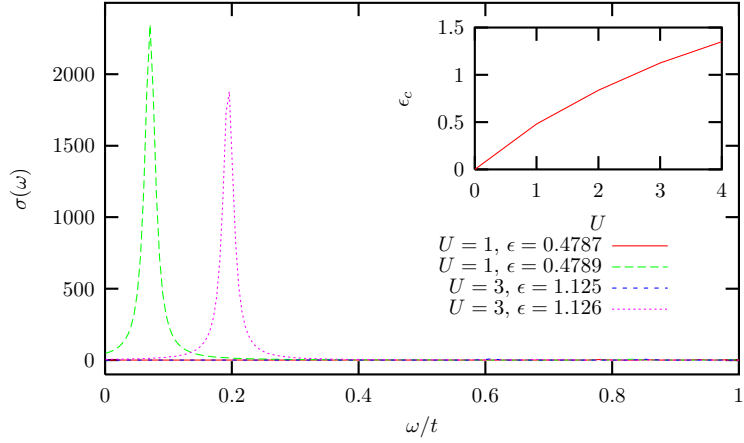}
\caption{Optical conductivity in arbitrary units versus frequency of the Hubbard model at
 third filling. System size and electron number are $L=9$ and $N_e=6$. The inset shows 
the characteristic impurity strength $\epsilon_c$, where the peak first arises,
versus interaction $U$.}
\label{fig-opt-Lutt}
\end{figure}

\begin{figure}[htb]
\includegraphics[width=0.6\textwidth]{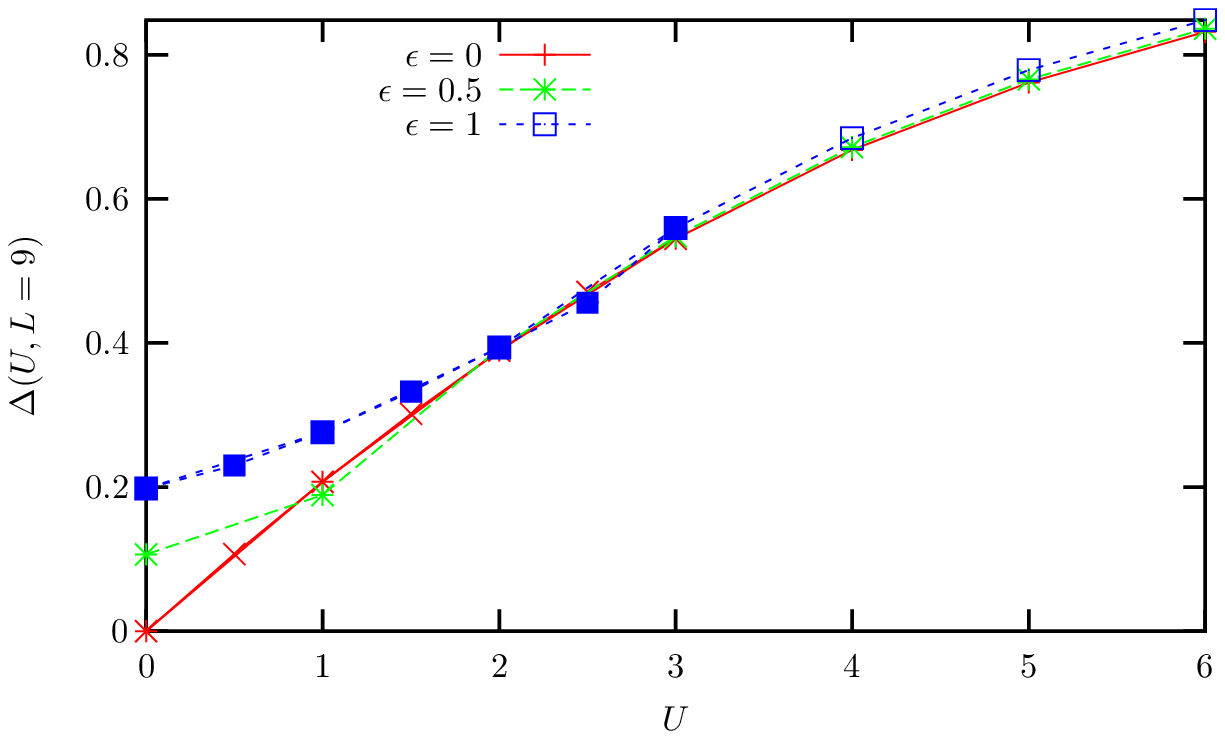}~a)\\
\includegraphics[width=0.6\textwidth]{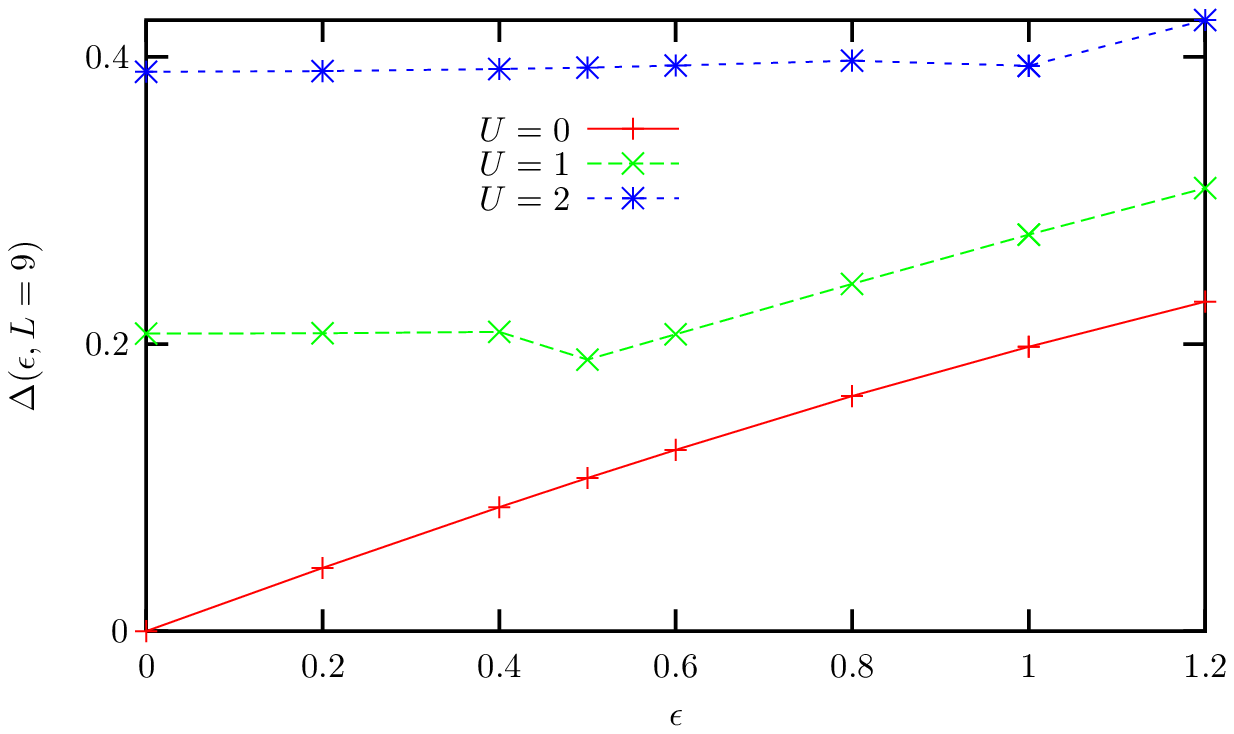}~b)
\caption{a) Energy gap versus interaction and fixed impurity strength.
b) Energy gap versus impurity strength  and fixed interaction.
In both cases, $L=9$ and $N_e=6$, and we use anti-periodic boundary conditions.} 
\label{fig-gap-Lutt}
\end{figure}

In the numerical data, 
the Drude peak occurs abruptly by increasing the impurity strength for fixed 
interaction strength, see Fig. \ref{fig-opt-Lutt}.
The transition is very sharp.
The characteristic impurity strength is proportional to the interaction,  
$\epsilon_c(U)\propto U/2$, as indicated in the inset of Fig. \ref{fig-opt-Lutt}.
By comparing with the energy gap, Eq. (\ref{gap-eq}), we see that the Drude peak occurs when
 $\Delta_L(U)$ becomes independent of impurity strength, or 
$\Delta_L(\epsilon)$ shows a dip at $\epsilon_c$, see Fig. \ref{fig-gap-Lutt}. 
Once raised, the peak position is proportional to the gap,
 $\omega_{\rm P}\propto \Delta(U,\epsilon)$.
We expect the Drude peak occurring for $2\epsilon \stackrel{ >}{\sim} U$ at 
$\omega_{\rm P}\approx 2\epsilon_{\rm eff}/L$ due to the discussion of the last subsection. With $\epsilon_{\rm eff}= \epsilon(L/L_0)^{1-(K_c+K_s)/2}$ \cite{Schm98}, we find
$\omega_{\rm P}\approx 2\epsilon L^{1-U/(2\pi v_F)}$.
We are thus able to follow in  the optical conductivity of   finite metallic systems the energy gap 
               on the basis the Drude peak. 
\subsection{Spin-gap phase}
\begin{figure}[htb]
\includegraphics[width=0.6\textwidth]{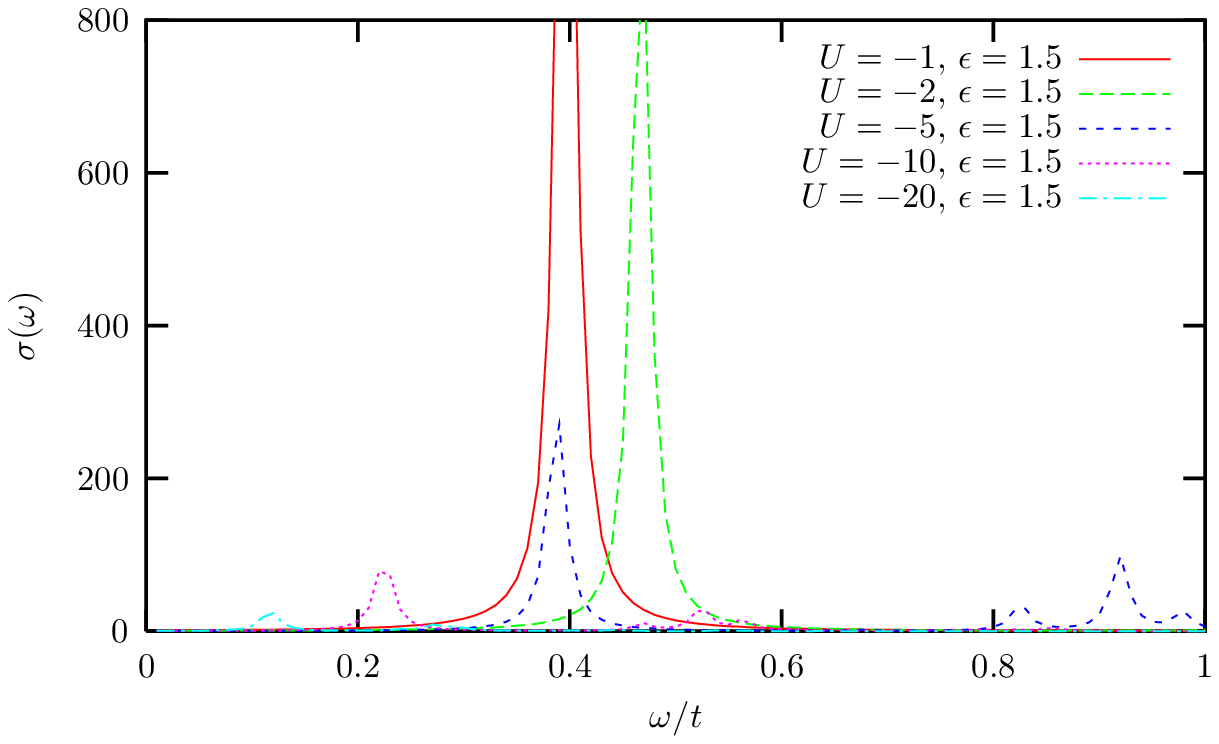}~a)\\
\includegraphics[width=0.6\textwidth]{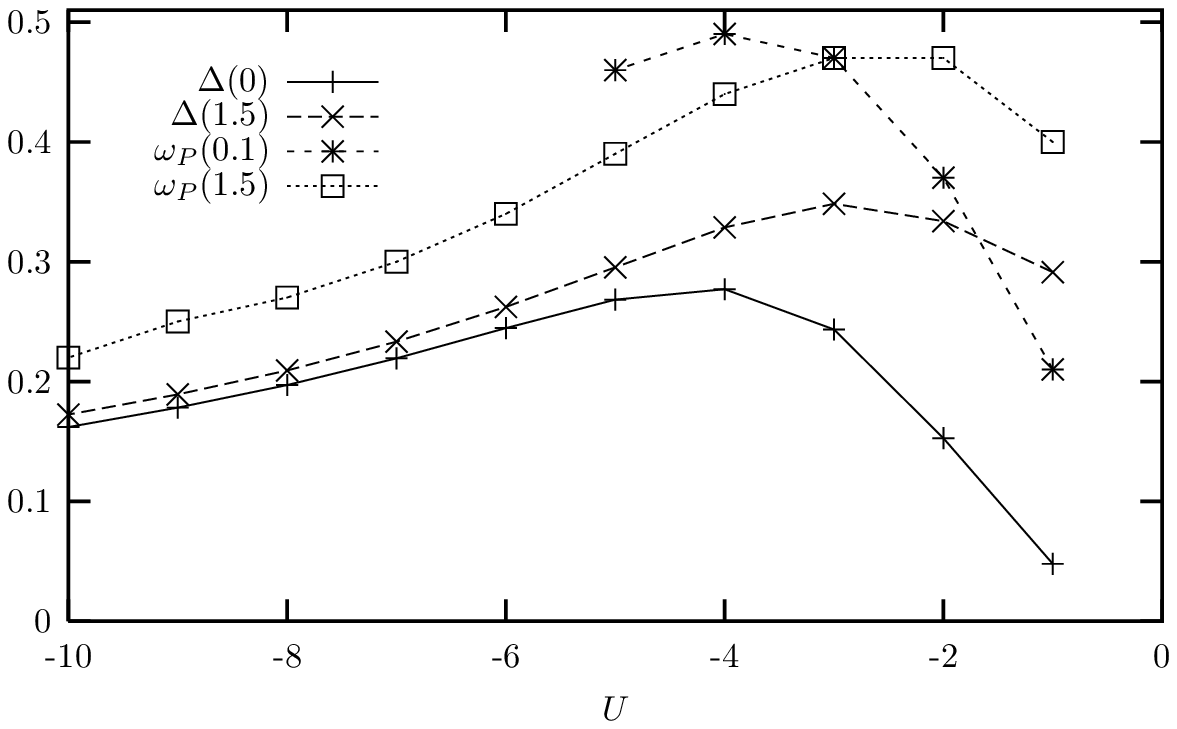}~b)
\caption{a) Optical conductivity in arbitrary units versus frequency. $L=10$, $N_e=10$.
 b) Energy gap and peak position -- taken from a) -- versus interaction. The lines are connecting the data points.}
\label{fig-spin-gap}
\end{figure}

Within the metallic phase for attractive interaction, where the
               spin excitations are  frozen out,
 the optical conductivity shows a similar behavior as in the Luttinger liquid phase. 
 Now, the superposition of the  open shell states is always the ground state.
In contrast to the Luttinger liquid phase, 
the Drude peak is present for {\em all} interaction and 
impurity strengths (in case of a {\em repulsive} impurity).  
The numerical results are shown in Fig. \ref{fig-spin-gap}a).
The peak position  marks again the energy gap, 
$\omega_{\rm P}\sim \Delta(U,\epsilon,L)$.  Both
               increase first with $|U|$, but decrease then
               with $1/|U|$, as shown in Fig. \ref{fig-spin-gap}b). 
               With periodic boundary
               conditions the gap decreases with  $1/U$-behavior over the whole parameter regime.
Due to the  spin gap, the peak loses drastically weight with increasing interaction strength.
\subsection{Mott insulating phase}
In case of half filling, the interaction lifts the 
degeneracy of the open shell states  and the partly filled shell states
\cite{Fye}. 
Again we refer to  the four site system with four electrons to write down the analytical expressions.
The fourfold degenerate open shell states with $E=-4t$ splits off in
 $E=-4t+3U/4$, $E=-4t+U$,   and
 $E=-4t+5U/4$, where
$E=\sum_n\varepsilon_n$. The second state with  $E=-4t+U$ is
  doubly degenerate and is represented by  
  the two spin-degenerate partly filled shell states shown in the upper left corner of Fig. \ref{fig-rand}.
The ground state with interaction is given by
the asymmetric superposition of the open shell states.  
For this reason, we see  in the numerical
 data for half filling a peak at  $\omega_P\approx \Delta(L) \approx U$, compare \cite{Jeckelmann00}.
 We will call 
 this peak `interaction' peak. 
In the insulator the question arises, whether the often supposed equality
between the  energy gap and the frequency of the interaction peak persists also in the
 system with impurity. 

\begin{figure}[htb]
\includegraphics[width=0.7\textwidth]{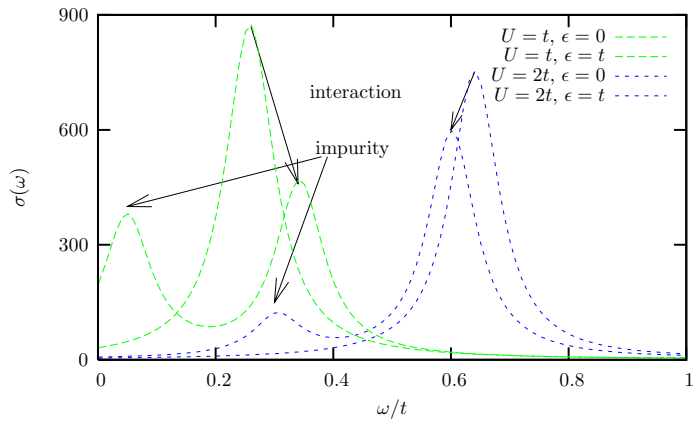}
\caption{ Optical conductivity in arbitrary units versus frequency. 
We use a smoothing of $\delta=0.1$.
System size and electron number is $L=10$ and $N_e=10$; we compare $U=t$ and $U=2t$ and
$\epsilon=0$ and $\epsilon=t$, respectively.} 
\label{fig-mott}
\end{figure}

 With weak impurity, compare the data for $U=2t$ and $\epsilon=t$ in  Fig. \ref{fig-mott},
 the position of this main peak shifts to lower frequencies as
the energy gap becomes smaller with increasing impurity strength.
An exception is the  case of small interaction, where
the gap decreases slightly with impurity strength, 
but the frequency of interaction peak
increases. For stronger interaction, both decrease with impurity strength.
Comparing the data taken from the optical conductivity with the energy gap, 
                it generally applies that interaction frequency and energy gap 
                agree only for weak impurities. 
In addition to the interaction peak another peaks grows at small frequencies due to the impurity.
The larger the interaction strength, the smaller is the weight of the impurity 
induced low-frequency peak.  On the other hand, the stronger the impurity, the smaller is the weight of the interaction peak.
\begin{figure}[htb]
\includegraphics[width=0.7\textwidth]{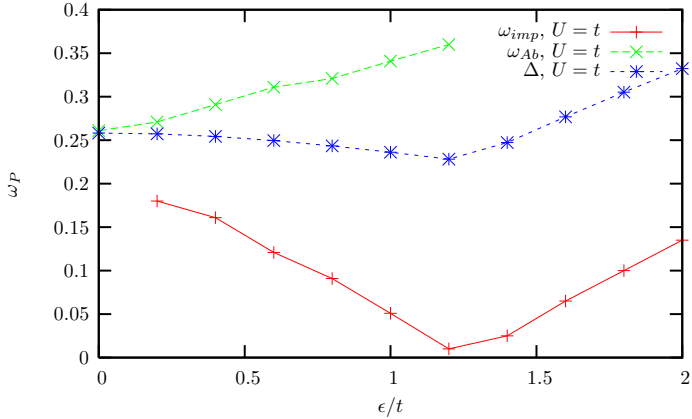}
\caption{Energy gap and peak positions, $\omega_{\rm Ab}$, for the interaction, and  $\omega_{\rm imp}$, for the impurity peak, 
versus impurity strength. The data are
 taken from Fig. \ref{fig-mott} for $U=t$.
The data for the energy gap are obtained again for $L=10$ and $N_e=10$ and anti-periodic boundary conditions. The lines are connecting the data points.}
\label{fig-mott2}
\end{figure}

In case of small interaction, compare $U=t$ and $\epsilon=t$ in  Fig. \ref{fig-mott},  or equivalently for 
strong impurities $\epsilon \approx U$,
the behavior is opposite. 
The spectrum of the optical conductivity is governed
  by the impurity.
The most surprising result is that the frequency of the impurity peak decreases to zero,
$\omega_{\rm imp}\to 0$ for $\epsilon \approx U$, see Fig. \ref{fig-mott2}.
At the same time, the peak gains weight.
Increasing the impurity strength further,
 only the impurity induced  peak survives and the interaction peak disappears. 
 Thus, within the short chains we can see that the impurity can
 shift the lower lying peak in the spectra to zero frequency,
 thereby indicating   a  metallic state. Nevertheless, the  
 energy gap is finite in this case. 
From this point of view,  the impurity is relevant in order to 
destroy the Mott insulating state. 
 Increasing the impurity strength further, 
 only the impurity induced  peak survives and the interaction peak is absent. 
 The impurity induced  peak then
  shifts to higher frequencies by further increasing the impurity strength -- comparable to the behavior in  the Luttinger phase. 

The results in the Mott insulating phase  clarify the results of our previous investigations, 
where we found a fast decay of the  Friedel oscillations  for small interaction \cite{Schreiber-rep}.
A similar behavior is known to exist for 
an irrelevant impurity in a metal.
The results concerning the phase sensitivity, which  show a slight maximum for 
small interaction in presence of a small impurity, indicates also a tendency to delocalization.
Due to
the strong influence of the impurity in the small gap regime, the correlation gap is closed and
gapless excitations are possible.  The resulting `metal' has other properties than the Luttinger 
liquid and we can find fast decaying Friedel oscillations.

\section{Summary and conclusions} 
In summary, a detailed determination of the ground state characteristics
of the one-dimensional Hubbard model in the presence of an impurity has
been achieved. In particular, we have
analyzed the optical conductivity of a small ring in all phases of the one-dimensional
Hubbard model.  
In the non-interacting system the Drude peak is seen
in small rings if appropriate 
boundary conditions are chosen, for which the ground state is degenerate.
Nevertheless, the continuation to the 
thermodynamic limit, where the transport properties should be independent from
the boundary conditions, is not clear.
Adding the impurity to the non-interacting system, 
the degeneracy of the ground state is lifted  and  
the optical conductivity 
shows  a clear peak at $\omega_{\rm P}
\sim \epsilon$. Thus, so to speak, the peak position maps 
the behavior of the finite size energy gap.  
A similar behavior is  found in the  Luttinger phase and the spin gap phase of the interacting system.

An impurity in the Mott insulating phase leads to an impurity state within the correlation gap.
The qualitative behavior of the optical conductivity is 
different for small and large -- compared to the interaction -- impurity strength.
 For large interaction the impurity has only a weak influence on the system, the interaction peak
 is only slightly shifted and has clearly more weight than the impurity peak.
On the other hand, in case of a  small energy gap, hence for small interaction,
  the  impurity dominates the behavior of $\sigma(\omega)$. The peak related to the impurity is only seen provided 
  $\epsilon > U$.
As in previous studies, we find that 
interaction and impurity -- both in the Luttinger and in the Mott phase --  weaken each other.

\section*{Acknowledgements}
We gratefully acknowledge various discussions with U. Eckern and D. Braak.
Financial
support  by the Deutsche Forschungsgemeinschaft within
Sonderforschungsbereich  484  and Schwerpunktprogramm 1073 is acknowledged.


\section*{References}
\begin{thebibliography}{10}
\bibitem{Carmelo}
J. M. P. Carmelo, N. M. R. Peres, and P. D. Sacramento
Phys. Rev. Lett. {\bf 84}, 4673 (2000)  
\bibitem{Fehske}
H.~Fehske, A.~P.~Kampf, M.~Sekania, and G.~Wellein, 
    Eur. Phys. J. B 31, 11-16 (2003);
    H. Fehske, G. Wellein, G. Hager, A. Weiße, and A.~R. Bishop,
Phys Rev. B \textbf{69}, 165115 (2004)
\bibitem{Mattis-Buch}
D.~C. Mattis, {\em The many-body problem}, (World Sientific, Singapore, 1993)
\bibitem{Shankar}
R. Shankar, Rev. Mod. Phys. 66, 129 (1994)
\bibitem{Schwerpunkt}
For a list of publications concerning one-dimensional correlated Fermi-systems see for example,
http://www.thp.uni-koeln.de/SP1073/veroeffentlichungen.html
\bibitem{Kramer-rep}
U.~Eckern and C.~Schuster,
{\it Quantum Coherence in Low-Dimensional 
       Interacting Fermi Systems},  edited by T. Brandes and S. Kettemann
(Springer-Verlag, Heidelberg, 2003).
\bibitem{Haldane81} F.~D.~M. Haldane, Phys. Rev. Lett. \textbf{47}, 1840 (1981)
\bibitem{Korepin} V.~E. Korepin, N.~M. Bogoliubov, and A.~G. Izergin,
{\em Quantum inverse scattering method and correlation functions}
(Cambridge University press, 1993)
\bibitem{Affleck}
I. Affleck, D. Gepner, H.~J. Schulz, and T. Ziman, J. Phys. A: Math.
Gen.  {\bf 22}, 511 (1989).
\bibitem{dmrg}
I.~Peschel, X.~Wang, M.~Kaulke, and K.~Hallberg,
\emph{Density-Matrix Renormalization:
A New Numerical Method in Physics}
(Lecture Notes in Physics, vol. 528, Springer, Heidelberg, 1999)
\bibitem{Dietmar}
 E. Gambetti-C\'{e}sare, D. Weinmann, R.~A. Jalabert, and P. Brune,
 Europhys. Lett. {\bf 60}, 120 (2002).
 \bibitem{Rudo}
 A.~W. Sandvik, D.~J. Scalapino, and P. Henelius, Phys. Rev. B {\bf 50}, 10474 (1994);
  R.~V. Pai, A. Punnoose,  and R.~A. R\"omer, arXiv:cond-mat/9704027
\bibitem{LiebWu}
E.~H. Lieb and F.~Y. Wu, Phys. Rev. Lett. \textbf{20}, 2435 (1968)
\bibitem{Yacoby}
Y. Tserkovnyak, B.~I.~Halperin, O.~M.~Ausl\"ander, and A. Yacoby,
Phys. Rev. B \textbf{68},  125312 (2003) 125312
\bibitem{Claessen}
R. Claessen, Physics World \textbf{15}(10), 22 (2002);
H. Benthien, F. Gebhard, and E. Jeckelmann,
Phys. Rev. Lett. {\bf 92}, 256401 (2004)
\bibitem{Voit}
J. Voit, Phys. Rev. B \textbf{45}, 4027 (1992)
\bibitem{Kane92}
C.~L. Kane and M.~P.~A. Fisher, Phys. Rev. Lett. \textbf{68},  1220  (1992)
\bibitem{Meden}
  V.~Meden, W.~Metzner, U.~Schollw\"ock, O.~Schneider, T.~Stauber, and K.~Sch\"onhammer,
  Eur.\ Phys.\ J.\ B {\bf 16},  631 (2000);
 V. Meden, W. Metzner, U. Schollw\"ock, and K. Sch\"onhammer, Phys. Rev. B
\textbf{65}, 045318 (2002).
\bibitem{Doty}
C.~A.~Doty and D.~S.~Fisher,  Phys. Rev. B \textbf{45}, 2167 (1992)
\bibitem{RomPun}
R. A. R\"omer and A. Punnoose, Phys. Rev. B \textbf{52}, 14809 (1995) 
\bibitem{mori}
H. Mori and Y. Takeuchi,
phys. stat. sol. (b) {\bf 241} (2004), 2189
\bibitem{B4G:Bed98}
  G.~Bed\"urftig, B.~Brendel, H.~Frahm, and R.~M.\ Noack,
  Phys.\ Rev.\ B {\bf 58}, 10225 (1998).
\bibitem{Schreiber-rep}
C.~Schuster and P.~Brune, phys. stat. sol. (b) \textbf{241} (2004), 2043
\bibitem{Rommer}
S. Rommer and S. Eggert, Phys. Rev. B \textbf{62},  4370 (2000)
Europhys. Lett. {\bf 60},  120 (2002)
\bibitem{Vollha}
 K. Byczuk, W. Hofstetter, and D. Vollhardt,
Phys. Rev. Lett. {\bf 94}, 056404 (2005)
\bibitem{Jeckelmann00}E.~Jeckelmann, F.~Gebhard, and F.~H.~L.~Essler,
  Phys.\ Rev.\ Lett.\ {\bf 85}, 3910 (2000);
E.~Jeckelmann, Phys.\ Rev. B \textbf{66}, 045115 (2002)
\bibitem{Schulz90}
H.~J. Schulz, Phys. Rev. Lett. \textbf{64}, 2831 (1990)
\bibitem{Shastry}
B.~S.~Shastry and B.~Sutherland, Phys. Rev. Lett. \textbf{65},
243 (1990) 
\bibitem{Leggett}
A.~J. Leggett, in {\em Granular Nanoelectrics}, edited by D.~K. Ferry (Plenum Press, New York, 1991), p. 297.  
\bibitem{Mattis}
D.~C. Mattis,  Phys. Rev. Lett. {\bf 32} (1974), 714.
\bibitem{Lanczos}
C. Lanczos, J. Res. Nat. Bur. Stand. {\bf 45}, 255 (1950).
\bibitem{Hallberg}
K.~A. Hallberg, Phys. Rev. B {\bf 52} (1995), 9827 
\bibitem{Kuehner}
  T.~D.\ K\"uhner and S.~R.\ White, Phys.\ Rev.\ B {\bf 60}, 335 (1999)
\bibitem{gagliano}
E.~R. Gagliano and C.~A. Balseiro, Phys. Rev. Lett. {\bf 59}, 2999 (1987)
\bibitem{Luther}
A. Luther and V.~J. Emery, Phys. Rev. Lett. {\bf 33} (1974), 589.
\bibitem{Schm98}
     P.~Schmitteckert, T.~Schulze, C.~Schuster, P.~Schwab, and U.~Eckern,
     Phys. Rev. Lett. \textbf{80}, 560 (1998)

\bibitem{Fye}
R.~M. Fye, M.~J. Martins, D.~J. Scalapino, J. Wagner, and W. Hanke,
Phys. Rev. B  {\bf 44} (1991), 6909
\end{thebibliography}
\end{document}